\documentclass[aps,prl,onecolumn,12 pt,superscriptaddress,showpacs]{revtex4}
\usepackage{amssymb}

\usepackage{graphicx}

\begin{document}

\title{Non-linear electronic transport and anomalous resistance fluctuations in the stripes state of $La_2NiO_{4.14}$}
\author{A. Pautrat}
\affiliation{Laboratoire CRISMAT, UMR 6508 du CNRS et de
l'ENSICAEN, 6 Bd Mar$\acute{e}$chal Juin, 14050 Caen, France.}
\author{F. Giovannelli}
\author{N. Poirot}
 \affiliation{Laboratoire LEMA, UMR 6157 CNRS-CEA,
Universit$\acute{e}$ F. Rabelais, Parc de Grandmont, 37200 Tours,
France.}

\begin{abstract}
We report on electronic transport measurements in
$La_2NiO_{4.14}$. Non-Ohmic and hysteretic $V(I)$ curves are
measured for $T \lesssim$ 220 $K$. Large and
non Gaussian resistance fluctuations can be observed, with strong cooling rate dependence. During a slow cooling, the
resistance reaches plateaus and then suddenly jumps for $T
\lesssim$ 100 $K$, evidencing a macroscopic freezing of the charges. Anti-correlation between time-series of orthogonal resistances is also observed.
These results are discussed in the framework of the stripes state
scenario.
\end{abstract}

\pacs{71.27.+a, 71.45.Lr, 72.20.-i, 72.20.Ht}

\newpage
\maketitle

In electronically doped Mott insulators, Coulomb interactions and
antiferromagnetic interactions between magnetic ions favor
localized charges, giving an insulating tendency. On the contrary,
the kinetic energy of doped holes tends to delocalize the charges.
The compromise has been shown to result in a charged stripes
state, where charge domains are between insulating
antiferromagnetic regions \cite{stripes}. This state is though to
be realized in layered nickelates and cuprates. Nevertheless,
their transport properties exhibit a major difference. When moderately doped with holes, cuprates can become
metal-like ("bad metal") and even superconducting, whereas
nickelates exhibit a semiconducting temperature variation up to
large hole doping. Extensive neutron scattering experiments have
given compelling evidence of the stripes scenario in
$La_{2-x}Sr_xNiO_{4+\delta}$ \cite{ishkawa}. Generally, "static"
stripes are observed, in agreement with the insulating nature of
the nickelates \cite{tranquada}. The stripes state can also be
modeled as an electronic liquid phase, using some analogy with
the liquid crystals \cite{liquid}. In particular, the conducting
properties of cuprates could be allowed by large transversal
fluctuations in a stripe liquid \cite{liquid}. Recent inelastic
neutron scattering experiments seem to rule out the relevance of
static stripes in $YBa_2Cu_3O_{7-{\delta}}$ (YBCO) \cite{philippe}, a canonical cuprate, but
leave a place for such fluctuating stripes \cite{vojta}.

Since a stripes state is a non uniform 2D electronic system,
unconventional transport properties are expected \cite{ando}. It is well known that non-linear transport is observed in charges
density waves (CDW) systems. CDW can be pinned by the quenched disorder
of the sample. Their depinning, above a threshold electric field,
leads to a decrease of the differential resistivity
\cite{monceau}. Equivalent features, i.e. the depinning and the associated collective motion, have been predicted 
for the stripes \cite{morais}. Interestingly, it allows
some analogies between the stripes dynamics problem and the more
general problem of elastic manifolds in quenched random media,
generally used for studying the vortex lattices and the CDW
dynamics. Unfortunately, the experimental situation is far from being clear in the stripes case. Transport non-linearities has
been observed in the charge ordered state of
$La_{2-x}Sr_xNiO_{4+\delta}$ \cite{tokura}, but overheating
effects can nuance the interpretation of the data
\cite{ando2,mercone}. The attempts to
measure non-linearities in the transport properties have been unsuccessful in $La_{2-x}Sr_{x}CuO_{4}$ , despite a detailed
study \cite{ando2}. A possible reason could be that the disorder is
not very effective to pin a liquid-like state. As a consequence,
the threshold electric field can be too large and not measurable. From another hand, resistance noise due to
fluctuations of charges domains can be expected \cite{carlson}, allowing a measure of pertinent properties even at low biasing current .
Recently, Bonetti \textit{et al} have observed telegraph-like resistance
jumps in the normal state of YBCO nanowires \cite{bonetti}, in
agreement with the existence of such fluctuating domains. In YBCO, the
existence of stripes is still controversial \cite{philippe}. It is thus important to know if
some related effects can be observed in a doped nickelate. This is one of the goal of
this experiment.

The sample is a single crystal of $La_2NiO_{4+\delta}$ with
dimensions of 4 $\times$ 1.9 $\times$ 0.5
 $mm^3$. The crystal was grown by image furnace. Our sample has an excess of oxygen corresponding to
$\delta =$ 0.14 $\pm$ 0.01, as determined by thermogravimetric
analysis. This is very close to the ideal oxygen excess of $\delta
=$ 2/15 of the interstitial structure \cite{oxyg,tran}. The
quality of the sample has been controlled by EDS, TEM and it was
oriented by Laue diffraction technique. The contacts were made
with a wire bonding system, connecting Aluminium wires on
evaporated gold pads separated by 150 $\mu m$, in a four probes
geometry.
 In a sample of the same composition, neutron scattering
studies have shown that charges order below an ill-defined
temperature $T \gtrsim$ 220 $K$. They have also evidenced that a clear
jump in the intensity of the charge superlattice peak occurs when cooling
through $T_{m} \approx$ 110 $K$, associated with the temperature of spin ordering
\cite{oxyg}. Several
 lock-in transitions of the wave vector of stripe order were also observed for
 $T \leq T_m$. The positions of the observed peaks indicate that the stripes run diagonally through the
 $NiO_{2}$ layers. In our geometry, this is perpendicular to the direction of the
 transport current (see inset Fig.1).

When dealing with transport non-linearities, a major problem is the overheating due to the Joule effect. In oxides,
the limiting points are the low thermal conductivity $\kappa_{th}$
($\kappa_{th}$ is about few $W/m/K$ for $T =$ 100-300 $K$), and the
semiconducting temperature variation of the resistance. A maximum transport current $I^{*}$ can be
applied before reaching a relevant Joule heating. When the applied
current is higher than $I^{*}$, a temperature gradient appears
between the sample middle and the surfaces, and a notable decrease
of the differential resistance is observed \cite{ando,mercone}.
Unfortunately, this effect mimics the depinning of a charge
ordered state. To be more quantitative, the $R(T)$ curve can be fitted by
$R=R_0 exp(-\beta (T/T_0-1))$, in order to have a simple analytic
solution when solving the thermal equations. In this case \cite{mercone}, $I^{*}=(2W) (2\kappa_{th} T_0 / \beta
\rho_0)^{1/2}$ . For $La_{2-x}Sr_{x}NiO_{4}$, $\kappa_{th}$
changes slightly from 300 to 60 $K$ (about
0.03 to 0.2 $W/K/cm$ for the $x=$ 0 compound, \cite{thermal}). The calculation leads to $I^{*}
\thickapprox$ 1 $mA$
 for $T_0=$ 60 $K$ ($\rho_0 =$ 54640 $\Omega.cm$, $\beta =$ 6.2), giving here $E^{*}\approx$ $10^4$ $V/cm$. This is close to the values reported in \cite{tokura}. In our sample, we observe a decrease of differential
 resistance ($dV/dI$) for values systematically lower than, but reasonably close to, this approximation (not shown). We conclude that the decrease of ($dV/dI$) at high current is likely a heating effect in our sample. Finally, to probe intrinsic processes, $I \lesssim$ 500 $\mu A$ was
employed for the $V(I)$ measurements, and all
the noise and time series measurements have been performed with a
"safe" current ($I \approx$ 1 $\mu A$ at high temperature and $I
\approx$ 50-500 $nA$ for $T \lesssim$ 110 $K$).

For $I \lesssim$ 1 $\mu A$, the resistance is found linear in the
temperature range $T =$ 60-300 $K$ (the lowest temperatures were
not studied due to the too large resistance of the sample).
The resistivity increases strongly when the temperature decreases, as previously reported \cite{poirot}, and
follows a functional form $R_0.exp (T^{*}/T)^{\mu}$ (Fig.1). From the
Zabrodskii plot, the plot of $ln(-\partial (ln R) / \partial
(ln T) )$ as function of $ln T$ , two different exponent $\mu$ are
evidenced (not shown): For 300 $K \gtrsim  T \gtrsim$ 110 $K$, $\mu=$ 0.85 $\pm$ 0.05
(close to a "soft gap"), and for 110 $K \gtrsim T \gtrsim$ 60 $K$,
$\mu=$ 0.59 $\pm$ 0.07 (close to a "variable range hopping" (VRH)).
In the "soft gap" regime, a small peak of the activation energy
can be observed at $T \approx$ 210 $K$. These exponents are
$\textit{not}$ in very good agreement with what is expected in the case of conventional VRH or
activated modeling. This is likely due to the intrinsic inhomogeneity
of the electronic ground state, which should invalidate conventional treatment. One can only note that the temperatures of
cross-over are close to the ordering temperatures extracted from
neutron scattering experiments \cite{oxyg}.

In Fig.2, the $V(I)$ curves are shown. They are measured at different
temperatures in the "safe" current regime previously discussed.
Both non-ohmicity and hysteresis are observed for $T \lesssim$ 220
$K$. This can be caused by a stripes depinning in the presence of
quenched disorder \cite{olson}. The threshold currents are in the
range 1-100 $\mu A$. This corresponds to rather low electric field
of 10$^{-1}$-10 $V/cm$, close to what is expected for a conventional CDW depinning \cite{monceau}. This
electric field exhibits an activated temperature dependence $E_{0}
exp(T^{*}/T)$, with energy barrier $T^{*} \approx$ 450 $K$ (not
shown). This is characteristic of a thermal assisted process, and
agrees also with some CDW depinning models \cite{monceau}. Anyway,
stripes or CDW depinning favors conductivity \cite{morais}, but the system becomes here $\textit{more}$ resistive at high current. After several cycling, we observed that the $V(I)$ curve
tends to be linear, meaning that the high resistance state is the
most stable one. We conclude that we do not observe a genuine
stripes depinning. A possibility is that the resistance switching
marks a transition to a stable state where
charges are more localized. The $V(I)$ measurements have been made after cooling the sample from 300 $K$ with a rate of about 0.5 $K/min$.
 A systematic cooling rate dependence of the $V(I)$ curves has not been performed.

We have shown that the $R(T)$, measured under a moderate
cooling rate condition (2 $K/min$), is smooth and without obvious
characteristics (see Fig.1). More insights into the existence of metastable states come from the study of resistance fluctuations. From an
experimental point of view, we have measured resistance
time-series $R(t)$, at a constant temperature after cooling the
sample, during long times (1-2 hours). From these time-series, the mean square value $\overline{(\delta R^*)^2}$ can be calculated
\cite{notebene}. The temperature variation of $\delta R^*$ is
shown in Fig.3. Typically, three regimes can be defined and will be discussed below. 

The first regime is for 240 $K \lesssim T \lesssim$ 300 $K$: the resistance is observed constant as function of the time whatever the cooling rate.
 The second regime is for 200 $K \lesssim T \lesssim$ 240 $K$. After a fast cooling of the sample from $T=$ 300
$K$, a slight and continuous decrease of the
resistance as a function of the time can be observed (not shown). After turning off the current during $1000
sec$ and then restoring it, the relaxation restarts as
if the current was not turned off. This shows that the system
has intrinsically evolved during the time. This effect can be the counterpart of the
very low kinetic of the growth of the interstitial oxygen ordering
after quenching the sample, such as reported from neutron
diffraction study \cite{ordering}. If the sample is cooled from $T =$ 300 $K$ to the same temperature, but with a low or moderate cooling rate ($\lesssim$ 1 $K/min$), the time serie is
smooth and stationary and does not contribute to the excess noise.

For $T \lesssim$ 200 $K$,  a third and more interesting regime is observed. $\delta R^*(T)$ exhibits large
peaks for several temperature values (Fig. 3). These peaks
correspond to non-stationary time-series with large resistance steps,
as evidenced in the inset of the Fig.3. The typical lifetime of
one resistance state is several hundred of seconds. After a very
long time (up to 2 hours), the resistance tends to a quasi-stationary behaviour, where only two-states switching can be
rarely evidenced. This is quite different than the
telegraphic-like noise observed in YBCO, this latter being
associated to random fluctuations of charges/stripes domains
\cite{bonetti}. In contrast, the non-stationarity observed here
implies a more complicated kinetic in $La_2NiO_{4.14}$. The
time-series look like crackling noise \cite{cracking}, i.e.
exhibit discrete events with different sizes, as shown in Fig.4.
 The histogram of the size of resistance steps $\Delta R$ has
power-law dependence $\Delta R \propto R^{-s}$ with $s \approx$
1.3, in general agreement with avalanche exponents found in other
systems \cite{vincent} (note that, in our case, the restricted
number of events does not allow to prove rigorously a power-law). For the
well documented case of magnetic avalanches,
the domain wall motion in a disordered landscape should be
responsible for the avalanche behavior \cite{barka}. In the
framework of stripes dynamics, this implies that disorder is
relevant in our doped nickelate, in agreement with \cite{hassel}.
We have also observed that the typical time scale of the resistance
fluctuations is dependent of the cooling rate, suggesting some
glassy dynamics. This effect is more pronounced at low temperature. In Fig. 5, the resistance
time-traces are shown at $T =$ 60 $K$, after a fast cooling ($5 K/min$) or a slow
cooling ($\lesssim$ 0.5 $K/min$) from $T =$ 300 $K$. We have noted that the general features are well reproduced after identical thermal histories.
The slow cooling
favors more probable resistance steps, proving that they do not
come from a frozen disorder due to temperature quenching. For such a
slow cooling, a very peculiar behavior can be observed. As shown in
Fig. 6, when measured continuously during the cooling, the resistance
tends to saturate and then suddenly jumps for some temperatures
$T \lesssim 100 K$. In the same temperature range, commensuration
effects have been observed between the stripes spacing and the
lattice spacing \cite{oxyg}. This shows strong coupling with the
lattice potential, and besides, sensitivity to disorder
\cite{hassel}. The plateaus in the $R(T)$ evidence a freezing of
the charges at a macroscopic scale due to this strong coupling. We
think that a slow cooling allows a strongest efficiency of the disorder landscape sampled by the system.

In addition, we have observed anticorrelation between resistance
fluctuations measured along and perpendicular to the current flow
(see Fig.7). This anticorrelation is particularly clear for the
large resistance steps. In \cite{carlson}, it has been
discussed that a nematic long range order, such as present in a
stripes phase, is preserved only locally when it is coupled with
the quenched disorder. The result is a 2D distribution of stripes
patches whose motion or reorganization can gives rise to crackling noise and to anticorrelation in resistance noise
\cite{carlson}. At higher temperatures ($T \gtrsim$ 100 $K$), this anticorrelation becomes unmeasurable. One can note that $T \approx$ 110 $K$
is a characteristic temperature in neutron scattering experiment, marked by a sudden and strong
increase of the neutron intensity, and associated with the stripes
ordering \cite{oxyg}. Our measurement of anticorrelation seems
consistent with a global picture of an ill-defined stripe order at high
temperature, which freezes, coupled to the quenched disorder, for
$T \lesssim$ 100 $K$.

To conclude, non-linear electronic transport and
anomalous resistance time-series have been measured in
$La_2NiO_{4.14}$. Unlike in YBCO where relatively quiet and two states
fluctuations have been measured \cite{bonetti}, strong non-stationary effects are observed in this nickelate. We do not observe the genuine
depinning of a charge ordered state, but rather a reorganization
of metastable states after cooling the sample. If the
link we make between these features and the stripes state is correct, this latter shows a
very slow and constrained kinetic, likely due to a strong coupling
with the sample disorder.

\begin{acknowledgments}
We acknowledge M. Rossel (RUCA) and E. Veron
(CRMHT) for respectively the TEM and MEB characterizations of the sample.
\end{acknowledgments}

\newpage
\vskip 2 cm
\begin{figure}[tbp]
\centering \includegraphics*[width=7cm]{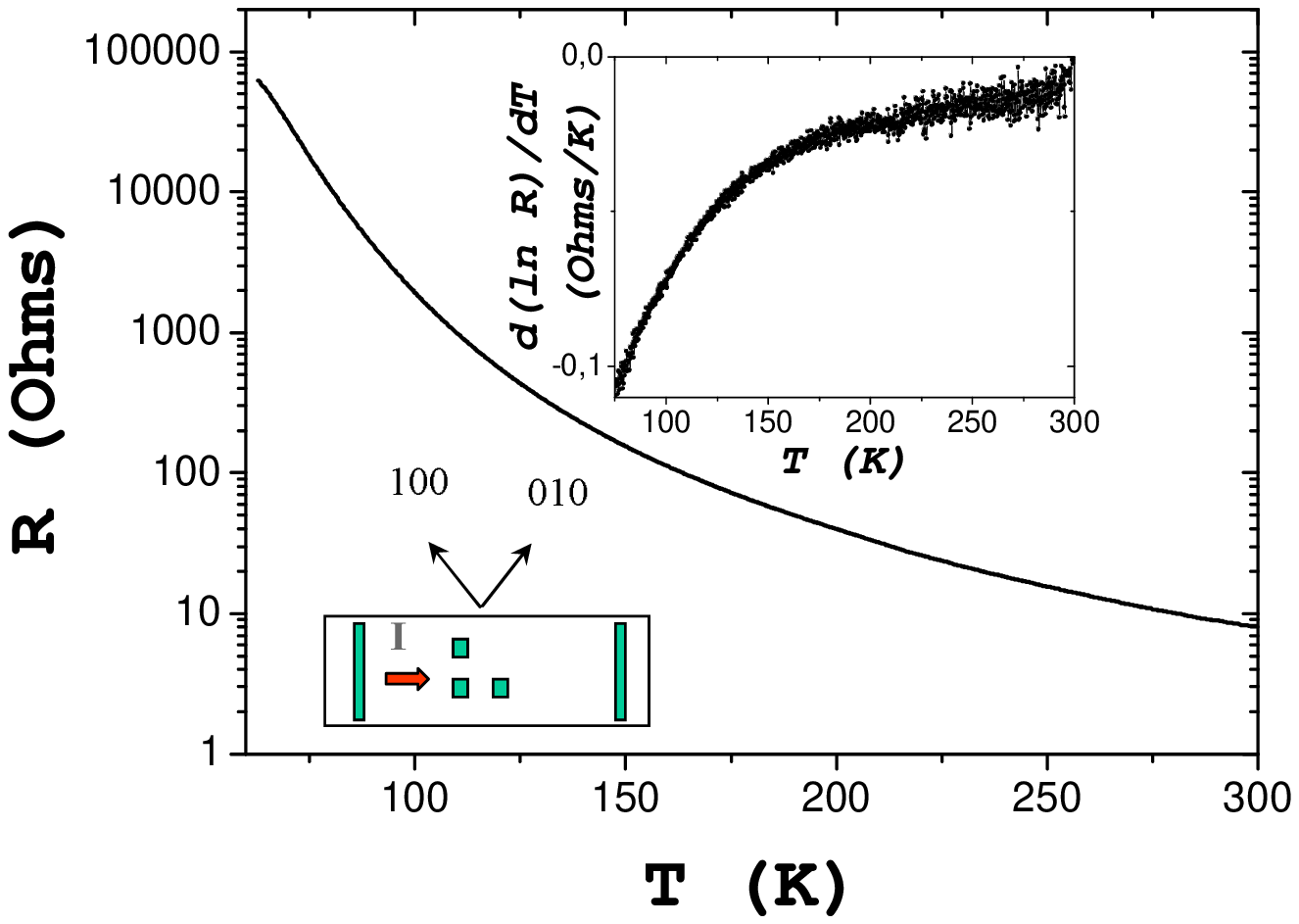}
\vskip 2 cm
\caption{(Color online) Low
current linear resistance of the $La_2NiO_{4.14}$ crystal as
function of the temperature  (cooling rate 2 $K/min$, $I=$ 1 $\mu A$). In the higher inset is shown the derivative of the resistance logarithm as a function of the temperature. 
 In the lower inset is shown the schematic view of the
sample with the contact pads. The cristallographic directions are also reported. $R$ stands for the
resistance along the transport current $I$, $R_{perp}$ is perpendicular to $I$.}
\end{figure}

\begin{figure}[tbp]
\centering \includegraphics*[width=7cm]{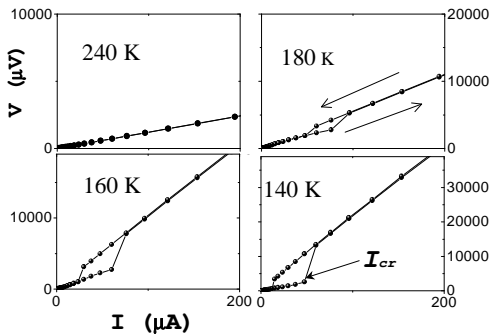} 
\vskip 2 cm
\caption{V(I)
curves for $T =$ 240, 180, 160 and 140 $K$. $I_{cr}$ corresponds
the current where a voltage jump can be observed when the
current is increased. The cooling rate before each acquisition is 0.5 $K/min$.}
\end{figure}

\begin{figure}[tbp]
\centering \includegraphics*[width=7cm]{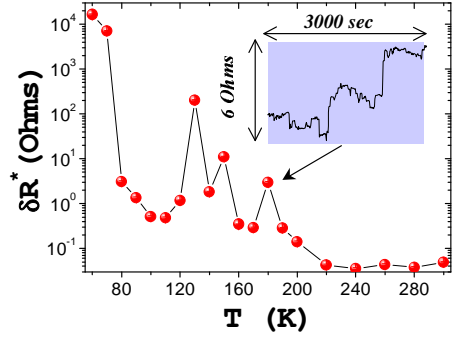}
\vskip 2 cm
\caption{(Color online) Resistance noise $\delta R^{*}$ (integrated over a
10$^{-2}$-10 $Hz$ bandwidth) as function of the temperature ($I =$ 1
$\mu A$). In the inset, a typical resistance time-serie is shown ($T
=$ 180 $K$). Between the noise peaks, the time-series are stationary
and without excess noise. The typical cooling rate is 0.5 $K/min$}
\end{figure}

\begin{figure}[tbp]
\centering \includegraphics*[width=7cm]{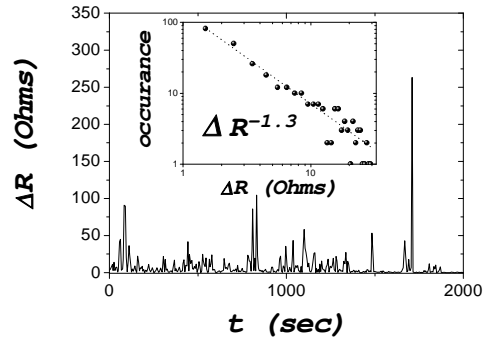}
\vskip 2 cm
\caption{Size
of the resistance steps $\Delta R$ as function of the acquisition time ($T =$
130 $K$, $I =$ 0.5 $\mu A$). In the inset is shown the histogram of
$\Delta R$ and the fit using a functional form $\Delta R^{-1.3}$.}
\end{figure}

\begin{figure}[tbp]
\centering \includegraphics*[width=7cm]{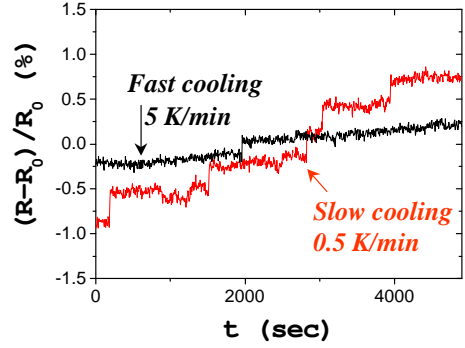}
\vskip 2 cm
\caption{(Color online) Time-traces of the resistance at $T =$ 60 $K$, after a slow
cooling or a fast cooling ($I =$  0.1 $\mu A$). Note
the increase of the rate of resistance steps after the slow
cooling.}
\end{figure}

\begin{figure}[tbp]
\centering \includegraphics*[width=7cm]{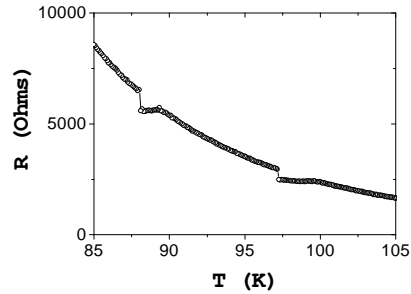}
\vskip 2 cm
\caption{Resistance as a function of the temperature during a slow
continuous cooling (rate= 0.5 $K/min$, $I =$  0.5 $\mu A$). Note the
two plateaus, which are followed by jumps to the "normal"
resistance value.}
\end{figure}

\begin{figure}[tbp]
\centering \includegraphics*[width=7cm]{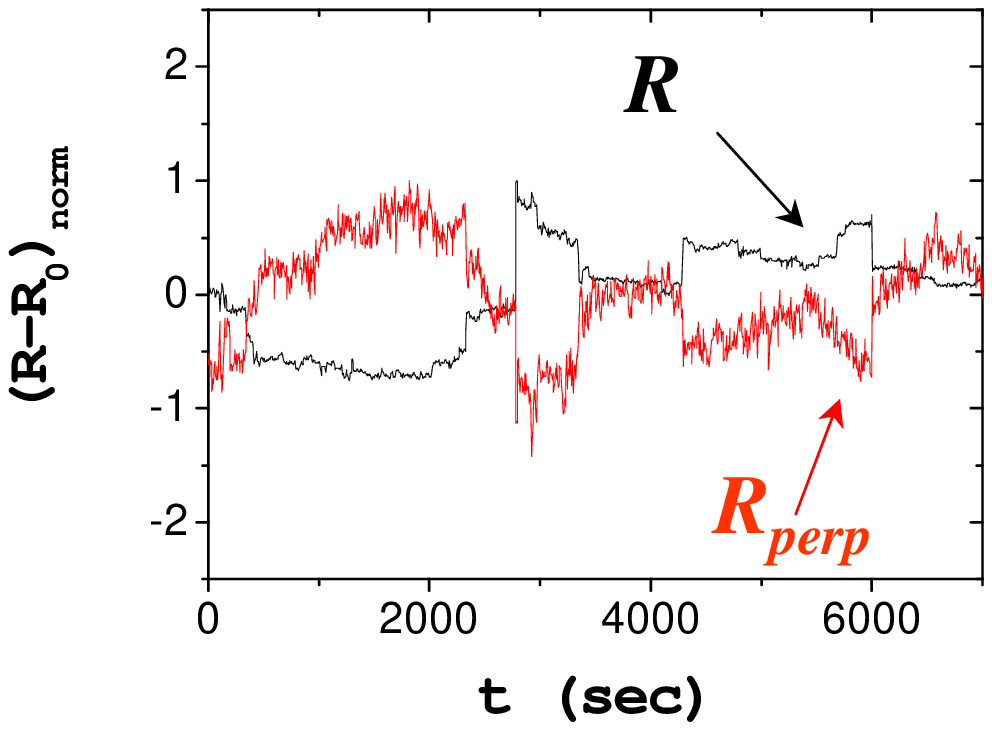}
\vskip 2 cm
 \caption{(Color online) Time
series of resistance in orthogonal directions (rate= 0.1 $K/min$, $T =$ 60 $K$, $I =$
1 $\mu A$). $R_{perp}$ is the resistance perpendicular to the direction
of the current. Each trace has been normalized. Strong anticorrelation can be observed between
the two time-series.}

\end{figure}


\begin{references}
\label{sec:TeXbooks}
\bibitem{stripes} V.J. Emery and V. J.  Kivelson, Physica C 209, 597
(1993).
\bibitem{ishkawa} N. Ichikawa, S. Uchida, J. M. Tranquada, T. Niemöller, P. M. Gehring, S.-H. Lee, and J. R. Schneider, Phys. Rev. Lett. 85, 1738 (2000).
\bibitem{tranquada} V. J. Emery, S. A. Kivelson, J. M. Tranquada, Proc. Natl. Acad. Sci. USA 96, 8814
(1999).
\bibitem{liquid}S. A. Kivelson, E. Fradkin, and V. J. Emery, Nature 393, 550 (1998).
\bibitem{philippe}  V. Hinkov, S. Pailhes, P. Bourges, Y.
Sidis, A. Ivanov, A. Kulakov, C.T. Lin, D.P. Chen, C. Bernhard,
and B. Keimer,  Nature 430, 650 (2004). V. Hinkov, P. Bourges, S.
Pailhes, Y. Sidis, A. Ivanov, C.T. Lin, D.P. Chen, B. Keimer,
cond-mat/0601048.
\bibitem{vojta} M. Vojta, T. Vojta, and R. K. Kaul, cond-mat/0510448
(2005).
\bibitem{ando} Y.Ando, A. N. Lavrov, S. Komiya, K. Segawa, and X. F.
Sun, Phys. Rev. Lett. 87, 017001 (2001).
\bibitem{monceau} G. Gr\"{u}ner, Rev. Mod. Phys. 60, 1129 (1988).
\bibitem{morais} C. Morais Smith, Yu. A. Dimashko, N. Hasselmann, and A. O. Caldeira
Phys. Rev. B 58, 453 (1998).
\bibitem{tokura} S. Yamanouchi, Y. Taguchi, and Y. Tokura, Phys. Rev. Lett. 83, 5555 (1999).
\bibitem{ando2} A. N. Lavrov, I. Tsukada, and Y. Ando, Phys. Rev. B 68, 094506
(2003).
\bibitem{mercone} Silvana Mercone, Raymond Fr\'{e}sard, Vincent Caignaert, Christine Martin, Damien Saurel,
 Charles Simon, Gilles Andr\'{e}, Philippe Monod, and Fran\c{c}ois
 Fauth, J. Appl. Phys. 98, 023911 (2005).
\bibitem{carlson} E. W. Carlson, K. A. Dahmen, E. Fradkin, and S. A.
Kivelson, Phys. Rev. Lett. 96, 097003 (2006).
\bibitem{bonetti} J. A. Bonetti, D. S. Caplan, D. J. Van Harlingen, and M. B.
Weissman, Phys. Rev. Lett. 93, 087002 (2004).
\bibitem{oxyg} P. Wochner, J. M. Tranquada, D. J. Buttrey, and V.
Sachan, Phys. Rev. B 57, 1066 (1998).
\bibitem{tran} J. M. Tranquada, J. E. Lorenzo, D. J. Buttrey, and V. Sachan, Phys. Rev. B 52, 3581 (1995).
\bibitem{thermal} J.Q. Yan, J.S. Zhou, and J. B. Goodenough, Phys.
Rev. B 68, 104520 (2003).
\bibitem{poirot} N. Poirot and F. Gervais, J. Superconductivity 18, 149 (2005).
\bibitem{olson} C. Reichhardt, C.J. Olson Reichhardt, and A.R. Bishop, Europhys. Lett. 72, 444 (2005).
\bibitem{notebene} $\delta R^*$ is here indicative of large fluctuations but does not have a rigorous meaning for non stationary
processes such as those observed here.
\bibitem{ordering} J.E. Lorenzo, J.M. Tranquada, D.J. Buttrey and
V. Sachan, Phys. Rev. B 51, 3176 (1995).
\bibitem{cracking}J.P. Sethna, K.A. Dahmen, and C.R. Myers, Nature 410, 242
(2001).
\bibitem{vincent} E. Vives, J. Goicoechea, J. Ortín, and A. Planes, Phys. Rev. E 52, R5 (1995).
\bibitem{barka} G. Durin and S. Zapperi, The Science of Hysteresis: Physical Modeling, Micromagnetics, and Magnetization
 Dynamics vol II (Amsterdam: Academic) chapter III (The Barkhausen Noise), 181 (2005)
 [cond-mat/0404512].
\bibitem{hassel} N. Hasselmann, A. H. Castro Neto, C. Morais Smith, and Y.
Dimashko, Phys. Rev. Lett. 82, 2135 (1999).

\end{references}
\end{document}